\documentclass[useAMS,usenatbib]{mn2e}
\usepackage{epsfig}
\usepackage{color}
\usepackage{subfigure}

\def\lsim{\mathrel{\rlap{\lower3.5pt\hbox{\hskip0.5pt$\sim$}}
    \raise0.5pt\hbox{$<$}}}
\def\gsim{~\rlap{$>$}{\lower 1.0ex\hbox{$\sim$}}}

\newcommand{\goodgap}{\hspace{\subfigtopskip} \hspace{\subfigbottomskip}}

\title[HL and spiral galaxies rotation curves]{Spiral galaxies rotation curves in the Horava\,-\,Lifshitz gravity theory}
\author[V.F. Cardone et al.]{V.F. Cardone$^{1}$, M. Capone$^{2,3}$, N. Radicella$^{4}$, M.L. Ruggiero$^{3,5}$\\
$^1$I.N.A.F. - Osservatorio Astronomico di Roma, via Frascati 33, 00040\,-\,Monte Porzio Catone (Roma), Italy \\
$^2$Dipartimento di Matematica, Universit\`{a} degli Studi di Torino, Via Carlo Alberto 10,  10125\,-\,Torino, Italy \\
$^3$I.N.F.N.\,-\,Sezione di Torino, via Pietro Giuria 1, 10125\,-\,Torino, Italy \\
$^4$Department de Fisica, Universitat Autonoma de Barcelona, 08193\,-\,Bellaterra (Barcelona), Spain \\
$^5$Dipartimento di Fisica, Politecnico di Torino, Corso Duca degli Abruzzi 24, 10129 - Torino, Italy \\}

\date{Accepted xxx, Received yyy, in original form zzz}

\begin{document}

\maketitle

\begin{abstract}

We focus on a modified version of Horava\,-\,Lifschitz theory and, in particular, we consider the impact of its weak\,-\,field static spherically symmetric limit on the galaxy dynamics. In a previous paper, we used the modified gravitational potential obtained in this theory to evaluate the Milky Way rotation curve using a spheroidal truncated power\,-\,law bulge and a double exponential disc as the only sources of the gravitational field and showed that the modified rotation curved is not in agreement with the data. Making a step forward, we here include also the contribution from a dark matter halo in order to see whether this helps fitting the rotation curve data. As a test case, we consider a sample of spiral galaxies with smooth baryon matter distribution and well measured circular velocity profiles. It turns out that, although a marginal agreement with the data can be found, this can only be obtained if the dark matter halo has an unrealistically small virial mass and incredibly large concentration. Such results can be interpreted as a strong evidence against the  reliability of  the  gravitational potential obtained in the modified version of Horava\,-\,Lifschitz theory that we consider.

\end{abstract}


\section{Introduction}

In an attempt to construct a theory of quantum gravity in 3 + 1 dimensions, decoupled from string theory, Horava proposed, in 2009, a theory which took advantage of theoretical concepts investigated in the field of condensed matter physics \citep{HLa}. Precisely, Horava's theory is analogue to a scalar\,-\,field model developed some years ago by Lifschitz in the framework of the theory of quantum critical phenomena, in which the full Lorentz symmetry emerges only at infrared (IR) fixed point. For this reason, one usually refers to it as the Horava\,-\,Lifschitz (HL) theory.  The main feature of this theory as a field quantum gravity model is the anisotropic scaling of the space and time dimensions. Because of this anisotropic scaling, one finds the time dimension plays a privileged role showing up in a reduced invariance: full diffeomorphism invariance is abandoned, while only a subset, foliation\,-\,preserving diffeomorphism, is kept. This foliation has the property of becoming coincident with the standard diffeomorphism invariance in the IR limit, where General Relativity (GR) is recovered \citep{HLb}. By the way, the properties causing the enthusiastic reception it has gained among physicists are certainly its being power\,-\,counting renormalizable and definitely finite \citep{HLc}. Many of its features, like its ultraviolet (UV) and IR behaviour \citep{UV IR}, its impact on cosmology \citep{cosm}, spherically symmetric solutions \citep{sph symm}, black holes \citep{BH} and Solar System tests \citep{harko,iorio2010,iorio2011} have been deeply analyzed. Unfortunately, its original formulation, based on two main assumptions, that is the {\it detailed balance condition} (that reduces the number of operators in the action and simplifies some properties of the quantum system) and the {\it projectability condition} (a certain part of the spacetime metric, the so-called lapse function, can be set globally to unity), soon showed lots of problems both in the IR limits \citep{probl} and intrinsic parity violation in the purely gravitational sector of the mode \citep{UV IR}. The attempt to fix those problems led to abandon the detailed balance condition, still retaining the projectability condition \citep{BH, KK, CaPo}.
\indent Although the discussion about the foundations and the possible phenomenological and conceptual implications of the HL theory and of its various modifications is still open (see e.g. \cite{sotiriou10} and references therein), we believe it is worth systematically investigating its consequences at every scales. Following this path, in a previous paper \citep{PapI} we have complemented recent works on the cosmological consequences of the modified HL gravity as presented in Sotiriou et al. (2009a,b) by addressing its impact on the gravitational potential\,: in contrast to the results obtained in Tang \& Chen (2009), we showed that static spherically symmetric solutions other than the Schwartzschild\,-\,de Sitter one do exist. Precisely, we worked out a general formalism to estimate the rotation curve (that is, the circular velocity, $v_{c}$, as a function of the distance from the centre, $R$) for an extended source, showing that the HL theory can boost the $v_{c}(R)$ with respect to the Newtonian value, as a consequence of the modification induced on the gravitational potential of a point mass. As an application, we evaluated the Milky Way rotation curve using a spheroidal truncated power-law bulge and a double exponential disc as the only sources of the gravitational field, finding that the modified rotation curve is unable to fit the data. This means that the modified HL model taken into account can not play the role of an alternative solution to the missing mass problem. Motivated by this conclusion, we therefore take here a step further adding a dark matter halo to the galaxy modelling. Provided the galaxy model is reliable, one should then be able to fit the rotation curve of spiral galaxies within the framework of the modified HL potential. Moreover, such a test would also allow us to constrain the modified potential parameters and hence put some constraints on the coupling quantities entering the HL Lagrangian. On the contrary, should the model provide a poor fit to the data, one can draw some interesting lessons on the validity of the HL gravity from a test performed on galactic scales thus complementing the theoretical investigations and the observational analyses based on cosmological data. It is interesting to point out that even though our work is based on a static spherically symmetric solution obtained in the framework of HL gravity, our results  can be applied  as well to an arbitrary perturbation of the Newtonian gravitational potential having the same form.

The plan of the paper is as follows. In Sect.\,2, we review the derivation of the rotation curve for extended systems in the framework of Horava\,-\,Lifschitz gravity and then apply it to a spiral galaxy modelled as the sum of a thin disc and spherical dark matter halo. The spiral galaxies sample and its main properties are described in Sect.\,3 together with the method used to constrain both the halo and theory parameters. The results are presented and discussed in Sect.\,4, while summary and conclusions are given in Sect.\,5.

\section{Horava\,-\,Lifschitz rotation curve}

As stated above, we consider here the version of the HL gravity proposed by  Sotiriou, Visser \& Weinfurtner (2009 a, b) whose weak field limit has been first discussed by Tang \& Chen (2010). In our previous paper \citep{PapI}, we extended the Tang \& Chen results by looking for a solution different than their Schwartzschild\,-\,de Sitter one. We indeed found that the gravitational potential generated by a point mass $m$ read\,:

\begin{equation}
\Phi(r) = \Phi_N(r) + \Phi_{HL}(r)\,,
\label{eq: phipoint}
\end{equation}
where $\Phi_N(r) = - Gm/r$ is the Newtonian potential and

\begin{equation}
\Phi_{HL}(r)=\frac{A}{12f}r^{2}+\frac{B}{4f}-\frac{D}{12f}\frac{1}{r^{4}}\,,
\label{eq:HL1point}
\end{equation}
is the correction due to the HL theory. Here, the $(A, B, D)$ parameters are related to the coupling constants entering the HL Lagrangian, while $f$ is an unknown function of the source mass $m$ satisfying the condition $f(m = 0) = 1$. The corrective term may be conveniently rewritten as \citep{PapI}\,:

\begin{equation}
\Phi_{HL}(r) = - \frac{G M_{\odot}}{r_s} \left [
- \left ( \frac{\eta}{\eta_A} \right )^2 - \frac{B r_s}{4 f G M_{\odot}}
+ \left ( \frac{\eta}{\eta_D} \right )^{-4} \right ]\,,
\label{eq: phihlpoint}
\end{equation}
with $M_{\odot}$ the Sun mass and $r_s$ an arbitrary chosen reference radius introduced to
define the dimensionless quantity $\eta = r/r_s$. In Eq.(\ref{eq: phihlpoint}), we have
finally defined the scaling radii\,:

\begin{equation}
\left \{
\begin{array}{lll}
\displaystyle{r_A} & = & \displaystyle{\left ( \frac{12 f_{\odot} G M_{\odot}}{A r_s} \right )^{1/2}}
\left [ \frac{f(\mu)}{f_{\odot}} \right ]^{1/2} \\
~ & ~ & ~ \\
\displaystyle{r_D} & = & \displaystyle{\left ( \frac{D_{\odot} r_s^{-3}}{12 f_{\odot} G M_{\odot}} \right )^{1/4}}
\left [ \frac{D(\mu)}{D_{\odot}} \frac{f_{\odot}}{f_(\mu)} \right ]^{1/4} \,,\\
\end{array}
\right .
\label{eq: defradiibis}
\end{equation}
where quantities labelled with a $\odot$ are evaluated for $\mu = m/M_{\odot} = 1$. It is worth noting that the second term in Eq.(\ref{eq: phihlpoint}) simply adds a constant to the potential which has no effect in any situation of interest, so that we will henceforth neglect this term. We stress that this by no way means that $B$ can be set to zero. Indeed, $B = 0$ means $f = 1$ so that also $D$ vanishes  (again, see \cite{PapI} for details) and we go back to the Schwartzschild\,-\,de Sitter solution, while we are here interested in the more general case. We therefore assume $B \neq 0$, but nevertheless hereafter neglect its contribution because it drops off from the derivation of the quantities we are interested in.

As Eq.(\ref{eq: phihlpoint}) refers to the case of a point source mass, we have first to generalize it to an extended system in order to derive the rotation curve $v_c(R) = R\, (\partial \Phi/\partial R)$. To this end, we have also to take care of the fact that the corrective term $\Phi_{HL}$ does depend on $m$ in a way that we do not explicitly know, so that a nonlinear dependence can not be excluded a priori. As such, the superposition principle does not hold anymore and a different procedure has to be worked out. This problem has been addressed in \cite{PapI}, where we derived a general expression which holds for any potential leading to a point mass gravitational force factorizable as\,:

\begin{displaymath}
F_p(\mu, r) = \frac{G M_{\odot}}{r_s^2} f_{\mu}(\mu) f_{r}(\eta)\,,
\end{displaymath}
with $f_{\mu}$ and $f_r(r)$ dimensionless functions depending on the particular form of the point mass gravitational potential $\Phi_p$. Remembering that ${\bf F}_p = - \nabla \Phi_p$, it is only a matter of algebra to show that\,:

\begin{displaymath}
f_{\mu}(\mu) = \mu \ \ , \ \ f_{r}(\eta) = 1/\eta^2 \ \ ,
\end{displaymath}
for the Newtonian potential, while they are\,:
\begin{displaymath}
f_{r} = \left ( \frac{2}{\eta_{A \odot}} \right ) \left ( \frac{\eta}{\eta_{A \odot}} \right ) \ \ ,
\ \ f_\mu = \left [ \frac{f(\mu)}{f_{\odot}} \right ]^{-1} \ \ ,
\end{displaymath}
\begin{displaymath}
f_{r} = \left ( \frac{4}{\eta_{D \odot}} \right ) \left ( \frac{\eta}{\eta_{D \odot}} \right )^{-5} \ \ , \ \
f_\mu = \frac{D(\mu)}{D_{\odot}} \frac{f_{\odot}}{f(\mu)} \ \ ,
\end{displaymath}
for the first and second term in $\Phi_{HL}$. Following the steps detailed in \cite{PapI}, one finally gets for the rotation curve the expression\,:

\begin{eqnarray}
v_c^2(R) & = & G \ \rho_0 \ R_d^2 \ \eta \nonumber \\
~ & \times & \frac{\int_{\mu_{min}}^{\mu_{max}}{f_{\mu}(\mu) \psi(\mu) d\mu}}
{\int_{\mu_{min}}^{\mu_{max}}{\psi(\mu) d\mu}} \nonumber \\
~ & \times & \int_{0}^{\infty}{\eta' d\eta' \int_{-\infty}^{\infty}{\tilde{\rho}(\eta', \zeta') d\zeta'
\int_{0}^{\pi}{f_{r}(\Delta_0) d\theta'}}}\,,
\label{eq: rotcurvegen}
\end{eqnarray}
with

\begin{equation}
\Delta_0 = \Delta(\theta = \zeta = 0) =
\left [ \eta^2 + \eta'^2 - 2 \eta \eta' \cos{\theta'} + \zeta'^2 \right ]^{1/2} \ ,
\label{eq: defdeltazero}
\end{equation}
and $\psi(\mu)$ mass function (hereafter MF) normalized so that

\begin{displaymath}
\int_{\mu_{min}}^{\mu_{max}}{\mu \psi(\mu) d\mu} = \rho_0/M_{\odot}^2\,,
\end{displaymath}
with $\rho_0$ the mass density in the neighborhood of the reference radius.

It is worth stressing that Eq.(\ref{eq: rotcurvegen}) is fully general and can be used to compute the rotation curve, provided the expression for $f_{\mu}(\mu)$ and $f_r(\eta)$ are given. As a consistency check, it is easy to show that, for the Newtonian potential, the term depending on the MF is identically equal to unity, so that Eq.(\ref{eq: rotcurvegen}) reduces to a simple rewriting of the standard result. For the HL term, we get a dependence on the MF through the multiplicative term on the second row of Eq.(\ref{eq: rotcurvegen}). Note that the MF here only plays the role of scaling up or down the rotation curve. This is a consequence of the HL corrective term (\ref{eq: phihlpoint}) not depending on $m$. As such, one has simply to sum the contribution of all the stars notwithstanding their mass and this is indeed what the MF term gives in the HL case. The total rotation curve for the HL theory will finally be given as\,:

\begin{displaymath}
v_c^2(R) = v_N^2(R) + v_{HL}^2(R) = v_N^2(R) + v_A^2(R) + v_D^2(R)\,,
\end{displaymath}
where one can use the results known in literature for the computation of the Newtonian term $v_N^2(R)$ and our general rule with $f_{\mu}(\mu)$ and $f_{r}(\eta)$ given above to estimate the HL contribution $v_{HL}^2(R)$. Note that, in order to evaluate these latter terms, we need to know not only the mass density (as in the Newtonian only case), but also the local MF. Since we do not know this function, we can not separately constrain the parameters $(r_{A \odot}, r_{D \odot})$. We can, however, easily skip this problem as follows. Let us consider the $v_A$ term and write it explicitly as\,:

\begin{eqnarray}
v_A^2(R, \eta_{A\odot}) & = & G \ \rho_0 \ R_d^2 \ \eta \ \left (2/\eta_{A\odot}^2 \right ) \nonumber \\
~ & \times & \frac{\int_{\mu_{\min}}^{\mu_{max}}{\left [ f(\mu)/f_{\odot} \right ]^{-1} \psi(\mu) d\mu}}{\int_{\mu_{min}}^{\mu_{max}}{\psi(\mu) d\mu}} \nonumber \\
~ & \times & \int_{0}^{\infty}{\eta' d\eta' \int_{-\infty}^{\infty}{\tilde{\rho}(\eta', \zeta') d\zeta'
\int_{0}^{\pi}{\Delta_0  d\theta'}}} \ .
\label{eq: vaetasun}
\end{eqnarray}
We can now define a new length scale $\eta_A$ as\,:

\begin{equation}
\eta_A = \eta_{A\odot} \left \{
\frac{\int_{\mu_{\min}}^{\mu_{max}}{\left [ f(\mu)/f_{\odot} \right ]^{-1} \psi(\mu) d\mu}}{\int_{\mu_{min}}^{\mu_{max}}{\psi(\mu) d\mu}}
\right \}^{-1/2}
\label{eq: etaadef}
\end{equation}
so that the first HL corrective term simply becomes\,:

\begin{eqnarray}
v_A^2(R, \eta_{A}) & = & G \ \rho_0 \ R_d^2 \ \eta \ \left (2/\eta_{A}^2 \right ) \nonumber \\
~ & \times & \int_{0}^{\infty}{\eta' d\eta' \int_{-\infty}^{\infty}{\tilde{\rho}(\eta', \zeta') d\zeta'
\int_{0}^{\pi}{\Delta_0  d\theta'}}}
\label{eq: vaetaend}
\end{eqnarray}
which can be evaluated, as function of $\eta_A$, even if the local MF is not known. Using a similar trick for the $v_D^2(R)$, we can now fit the model to the rotation curve data thus constraining the parameters $(\eta_A, \eta_D)$. Should the local MF be known and a theoretically motivated expression for $f(\mu)$ be available, one can then translate the estimates of $(\eta_A, \eta_D)$ into constraints on $(\eta_{A \odot}, \eta_{D \odot})$.

Eq.(\ref{eq: vaetaend}) makes it clear that the shape of the local MF and the functional expression of $f(\mu)$ affects the overall shape of the total rotation curve through a rescaling of the corrective terms scalelengths $(\eta_A, \eta_D)$. On the other hand, it is possible that one has to set these latter quantities on a case\,-\,by\,-\,case basis in order to fit the rotation curves data. Actually, this is not a problem for the consistency of the theory since $(\eta_A, \eta_D)$ should not be universal quantities being related to the HL Lagrangian couplings only through integrals involving both $f(\mu)$ and $\psi(\mu)$. However, since we expect $f(\mu)$ to be the same for all systems, one has to assume that the local MF is different from one system to another in order to make a non universality of $(\eta_A R_d, \eta_D R_d)$ consistent with the constancy of $(r_A, r_D)$. Actually, whether the MF is universal or not is still an open question which we will not address here.

\section{Spiral galaxies rotation curves}

Historically, the flatness of spiral galaxies rotation curves was the first and (for a long time) more convincing evidence for the existence of dark matter \citep{SR01}. It is therefore a mandatory test for a whatever modified gravity theory to first demonstrate it is able to reproduce the observed circular velocity profile. Some care is, however, needed to be sure that the test results are not biased by possible systematics in both the theoretical prediction and the observational measurements. First, it is worth stressing that Eq.(\ref{eq: rotcurvegen}) implicitly assumes that the main contribution to $v_c(R)$ comes from stars moving on circular orbits in the disk plane. On the contrary, the asymmetric drift and the contribution from stars moving in the spiral arms may lead to non circular motions which are not taken into account in the above derivation. Moreover, since the HL corrective terms have to be computed numerically, a clumpy matter distribution can introduce an artificial bias related to how the density profile has been sampled and interpolated. In order to avoid both these problems, we therefore have to select spiral galaxies dominated by the stellar component and with a small contribution from the spiral arms such as Sbc, Sc, Sd systems, checking that the disk surface brightness profile can indeed be well approximated by the exponential model we have considered above.

From an observational point of view, the desiderata for the rotation curve data are mainly dictated by the need to sample the circular velocity profile with both high precision (so that significant constraints can be derived) and over a large radial range (in order to probe deep into the outer halo dominated regions). It is also highly preferable that the disk inclination has been well determined so that no systematic bias is induced on the measured $v_c(R)$ values. Finally, it is worth remembering that what one actually measures is the circular velocity as a function of the radial distance from the galaxy centre measured in arcsec. In order to convert to the linear units to get $v_c = v_c(r)$ with $r$ in kpc, we need to know wit great accuracy the galaxy distance from us. An uncertainty on this quantity translates in a global rescaling of the mass related parameters thus making difficult to raw any definitive conclusion on the importance of the dark matter contribution to the total mass budget.

Fulfilling all the above constraints is not an easy task and greatly reduce the sample of candidates for our analysis. Actually, it is worth stressing that our aim here is not to constrain the theory parameters, but rather to check its overall viability. From this point of view, a single Sc galaxy with an exponential disk and rotation curve data well sampled out to large radius is preferable to a large sample of systems with poorly determined rotation curves and dominated by not smooth disks. Indeed, should the HL modified potential fail to reproduce the data for this golden probe, one could safely argue against the theory as a whole.

\subsection{The data}

Motivated by the above considerations, we have looked at the THINGS \citep{Things} dataset comprising 19 galaxies with high quality rotation curves data finely sampling a large radial range (up to $\sim 10 R_d$). Both the inclination and the distance of these systems have been determined with high precision, while a careful analysis of the non circular motions have shown that these latter are negligible for most of the cases. The THINGS galaxies are, therefore, ideal candidates for our analysis fulfilling all the observational requirements we have hinted at in the above discussion.

Notwithstanding their high quality circular velocity measurements, not all the galaxies in the THINGS sample can be used for our analysis. Indeed, in order to not introduce spurious errors in the the theoretical rotation curve derivation, we ask that the stellar components have a smooth profile. To this regard, it is worth remembering that the luminosity profiles of the THINGS galaxies are typically described as the sum of an exponential disk and a second inner disk referred to as the bulge (only present for some galaxies). The mass density profile is then obtained multiplying the surface brightness by a stellar $M/L$ ratio inferred from the galaxy color. Since the color can radially change, the $M/L$ ratio could also not be constant so that the final density can be significantly different from the classical exponential profile and present undesirable wiggles and ripples. In order to avoid possible problems with the interpolation of non smooth components, we therefore select only galaxies with a smooth component and set the total disk mass $M_d$ and scalelength radius $R_d$ in such a way that the Newtonian rotation curve reported in the THINGS data is the same as the one of a Freeman (1970) disk (see later) with the same $(M_d, R_d)$ values. Such a smoothness cut gives a sample of only six galaxies. After excluding two systems with evidences for strong irregularities in the circular velocity profile, we select the four galaxies in Table\,\ref{tab: galdata} briefly presented below.

\begin{table}
\caption{Properties of sample galaxies. Explanation of the columns\,: name of the galaxy; logarithm of the bulge total mass (in solar units); bulge scalelength (in kpc); logarithm of the disk total mass (in solar units); disk scalelength (in kpc). Values for the bulge are not reported if there is no evidence for such a component in the surface brightness profile.}
\begin{center}
\begin{tabular}{|c|c|c|c|c|}
\hline
Id & $\log{M_b}$ & $R_b$ & $\log{M_d}$ & $R_d$ \\
\hline
NGC 2841 & 10.41 & 0.67 & 11.04 & 4.13 \\
NGC 3621 & ----- & ---- & 10.24 & 2.20 \\
NGC 5055 & 9.33 & 0.53 & 11.08 & 3.71 \\
NGC 6946 & 9.55 & 0.45 & 10.78 & 3.24 \\
\hline
\end{tabular}
\end{center}
\label{tab: galdata}
\end{table}

\begin{itemize}

\item{{\it NGC\,2841.} This is an early\,-\,type (Sb) spiral with absolute $B$ mag $M_B = -21.21$ and distance $D = 14.1 \ {\rm Mpc}$. Its rotation curve was first measured by \cite{B87} and shows a good overall agreement with the THINGS data we use here. Similarly, the position angle and the inclination differ by only few degrees from $P.A. = 153^{o}$ and $i = 74^o$ recommended by \cite{Things}.} \\

\item{{\it NGC\,3621.} At a distance $D = 6.6 \ {\rm Mpc}$, NGC\,3621 is a late\,-\,type spiral with a regular HI distribution. Although it is quite luminous with $M_B = -20.06$, no other determination of the rotation curve was available before the THINGS one which have also determine the position angle and inclination as $345^o$ and $65^o$, respectively.} \\
    
\item{{\it NGC\,5055.} This Sbc galaxy ($D = 10.1 \ {\rm Mpc}$, $M_B = -21.12$) has an extended and warped tenuous outer HI disc which originates the wiggle visible for large $R$ values in its rotation curve. The THINGS rotation curve agrees well with the previous determination \citep{B06} and extends further out thus being able to probe the warp for the first time. Position angle $(P.A. = 102^o$) and inclination $(i = 59^o)$ are also in agreement.} \\

\item{{\it NGC\,6946.} Although well known for its population of HI structures, the dynamics of NGC\,6946, a $M_B = -20.61$ late\,-\,type spiral at $D = 5.9 \ {\rm Mpc}$, has been relatively little studied because of its low inclination ($i = 33^o$, $P.A. = 243^o$). The offset between the THINGS data we use here and previous determinations \citep{C90,Boo07} is likely due to different assumptions for $i$.} \\

\end{itemize}
Although quite limited in number, such a small sample is nevertheless ideal for our test of the HL potential. Indeed, the regularity of the mass components, the detailed sampling of the rotation curve, the unambiguous determination of both the inclination and the distance leave almost no space for questioning the validity of the fit results. Indeed, should the fit work bad, one can confidently argue that it is a failure of the theory and not an outcome of some unaccounted for theoretical or observational bias. Motivated by this consideration, we prefer to not relax our severe selection criteria thus losing statistics, but improving the reliability of any inference on the validity of the HL modified potential.

\subsection{Modelling spiral galaxies}

In \cite{PapI}, we have attempted to fit the Milky Way rotation curve using the modified HL potential and the luminous disk, only in order to see whether it is possible to give off the dark matter contribution still obtaining a flat circular velocity profile. Such a test unambiguously demonstrated that this is not the case, so that a dark halo is still unavoidable to fit the rotation curve data. Actually, such a result could be anticipated looking at the scaling with $r$ of the additive terms in the point mass gravitational potential. Indeed, while the one $\propto r^{-4}$ quickly decreases, the correction scaling as $r^2$ should be weighted by a very large $r_A$ in order to avoid being dominant in the inner regions where the rotation curve is consistent with the prediction from Newtonian gravity. On the contrary, $r_A$ must not be too large if one wants to boost the circular velocity in the usually dark matter dominated region. Balancing these two opposite desiderata is actually not possible, so that a dark halo is still needed to fit the data. We will therefore model spiral galaxies as two component systems\footnote{Note that we are here neglecting the gas component since, for the galaxies considered, its contribution to the mass budget is much smaller than both the stars and the dark matter.}, namely the stars (in the bulge and the disk) and the dark matter.

We assume the stars (both for the bulge and the disk) are distributed in an infinitely thin and circularly symmetric disk. The surface density profile simply reads\,:

\begin{displaymath}
\Sigma(R) = \frac{M_d}{2 \pi R_d^2} \exp{\left ( -R/R_d \right )} \ ,
\end{displaymath}
with $(M_d, R_d)$ the total mass and the scalelength radius. The Newtonian rotation curve is given by \citep{Freeman70}\,:

\begin{eqnarray}
v_{dN}^2(R) & = & 2 \pi G \Sigma_0 R_d (\eta/2)^2  \nonumber \\
~ & {\times} & \left [ I_0(\eta/2) K_0(\eta/2) - I_1(\eta/2) K_1(\eta/2) \right ]\,,
\label{eq: vcdisknewt}
\end{eqnarray}
with $\eta = R/R_d$ and $I_l, K_l$ Bessel functions of order $l$ of the first and second type, respectively.

While the observed photometry motivates the use of the exponential profile for the disk, the choice of the dark halo model is not trivial.
Numerical simulations of structure formation are typically invoked as a direct evidence favouring the use of the NFW density law \citep{NFW96, NFW97} or its variants \citep{Moore+98, JS00}. However, the NFW model is the outcome of DM only for simulations performed in a Newtonian framework, while here we are working in a modified gravity theory. In principle, one should therefore rely on the results of simulations which include both the effect of the different potential and the impact on the evolution of structures due to deviations from GR. While this has been done in some cases, for instance in the context of $f(R)$ theories \citep{S09}, we are not aware of any similar test for HL gravity. We can nevertheless make some qualitative arguments to drive the choice of the halo profile. Firstly, we note that, in the intermediate regions, it is reasonable to expect that the $r^{-4}$ corrective term has yet faded away, while the $r^2$ one is still too small if $r_A$ is large enough. In this same limit, we can also extrapolate that, close to the virial radius, the potential is still approximately Newtonian, so that the results from N\,-\,body simulations can be taken as a first order approximation. On the contrary, in the inner regions, the $r^{-4}$ term can give a significant contribution, so that one can not rely on the scaling with $r$ inferred from Newtonian simulations, but a different density profile could be expected. An efficient way to parameterize these qualitative considerations is represented by the generalized NFW model \citep{JS00} reading\,:

\begin{equation}
\rho_h(r) = \frac{M_{vir}}{4 \pi R_s^3 g(R_{vir}/R_s)} \left ( \frac{r}{R_s} \right )^{-\alpha} \left ( 1 + \frac{r}{R_s} \right )^{-(3 - \alpha)}\,,
\label{eq: rhohalo}
\end{equation}
with

\begin{equation}
g(x) = (-1)^{\alpha} B(x, 3 - \alpha, \alpha - 2)\,,
\label{eq: defgx}
\end{equation}
and $B(x, a, b)$ the Beta function. In Eq.(\ref{eq: rhohalo}), $M_{vir}$ and $R_{vir}$ are the virial mass and radius, respectively. They are not independent, as the following relation holds

\begin{displaymath}
R_{vir} = \left ( \frac{3 M_{vir}}{4 \pi \Delta_{th} \bar{\rho}_M} \right )^{1/3}\,,
\end{displaymath}
with $\Delta_{th}$ the overdensity for spherical collapse and $\bar{\rho}_M = 3 H_0^2 \Omega_M/8 \pi G$ the mean matter density today. We follow \cite{BN98} for $\Delta_{th}$ and set $(\Omega_M, h) = (0.28, 0.70)$ in accordance with \cite{WMAP7}. Note that the value of $\Delta_{th}$ has been derived for a concordance $\Lambda$CDM model, hence implicitly assuming that GR rather than HL is the theory of gravity. However, from the cosmological point of view, HL reduces to $\Lambda$CDM for a spatially flat homogeneous and isotropic universe, so that no bias is induced by this assumption. Because of the spherical symmetry, the Newtonian rotation curve may then be easily evaluated as\,:

\begin{equation}
v_{hN}^2(r) = \frac{G M_h(r)}{r} = \frac{G M_{vir}}{R_{vir}} \frac{g(r/R_s)}{g(R_{vir}/R_s)} \ ,
\label{eq: vchalonewt}
\end{equation}
while the HL corrective term will be computed numerically. Note that, to this end, we need to assign the three halo parameters which we choose to be the inner regions logarithmic density slope, $\alpha$, the circular velocity at the virial radius, $V_{vir}^2 = G M_{vir}/R_{vir}$, and the concentration $c_{vir} = R_{vir}/R_s$. Note that we prefer to use $(V_{vir}, c_{vir})$ instead of $(M_{vir}, R_s)$ as fitting parameters since these are better constrained by the data being related, respectively, to the overall scale of the rotation curve and the balance between the stellar and dark matter terms in the intermediate radial regions.

\begin{table*}
\caption{Best fit solutions and marginalized constraints on the model parameters. Columns are as follows\,: 1. galaxy id, 2. best fit parameters, 3., 4., 5., 6., 7., 8. median and $68\%$ confidence ranges for the fitted quantities.}
\begin{center}
\begin{tabular}{|c|c|c|c|c|c|c|c|}
\hline
Id & ${\rm p}_{bf}$ & $\Upsilon/\Upsilon_{fid}$ & $\alpha$ & $c_{vir}$ & $V_{vir}$ & $\log{\eta_A}$ & $\log{\eta_D}$ \\
\hline

~ & ~ & ~ & ~ & ~ & ~ & ~ & ~ \\

NGC 2841 & (1.82, 0.21, 731, 88, 2.95, -0.65) & $1.93_{-0.19}^{+0.21}$ & $0.15_{-0.11}^{+0.37}$ & $396_{-223}^{+328}$ & $51_{-21}^{+32}$ & $1.61_{-0.47}^{+0.84}$ & $-0.66_{-0.58}^{+0.35}$ \\

~ & ~ & ~ & ~ & ~ & ~ & ~ & ~ \\

NGC 3621 & (1.05, 0.08, 211, 25, 1.67, -1.11) & $1.14_{-0.17}^{+0.23}$ & $0.17_{-0.12}^{+0.28}$ & $291_{-190}^{+330}$ & $25_{-14}^{+14}$ & $0.63_{-0.17}^{+0.39}$ & $-0.84_{-0.44}^{+0.25}$ \\

~ & ~ & ~ & ~ & ~ & ~ & ~ & ~ \\

NGC 5055 & (0.71, 0.08, 746, 75, 2.32, -0.38) & $0.79_{-0.09}^{+0.09}$ & $0.20_{-0.14}^{+0.27}$ & $168_{-80}^{+388}$ & $25_{-13}^{+27}$ & $1.14_{-0.27}^{+1.05}$ & $-0.60_{-0.74}^{+0.16}$ \\

~ & ~ & ~ & ~ & ~ & ~ & ~ & ~ \\

NGC 6946 & (0.78, 0.24, 16, 7, 0.96, -1.10) & $0.80_{-0.28}^{+0.20}$ & $0.17_{-0.12}^{+0.28}$ & $267_{-203}^{+323}$ & $26_{-16}^{+31}$ & $0.72_{-0.15}^{+0.59}$ & $-0.77_{-0.61}^{+0.31}$ \\

~ & ~ & ~ & ~ & ~ & ~ & ~ & ~ \\

\hline
\end{tabular}
\end{center}
\end{table*}

\subsection{Fitting procedure}

In order to constrain both the HL and halo parameters, we employ a standard Bayesian approach, first defining the likelihood function as\,:

\begin{eqnarray}
{\cal{L}}({\bf p}) & \propto & \exp{\left [ - \frac{\chi^2({\bf p})}{2} \right ]} \nonumber \\
~ & = & \exp{\left \{ - \frac{1}{2} \sum{\left [ \frac{v_c^{obs}(R_i) - v_c^{th}(R_i)}{\varepsilon_i}
\right ]^2} \right \}}\,,
\label{eq: deflike}
\end{eqnarray}
where ${\bf p}$ is the set of the parameters of the model, $v_c^{obs}(R_i)$ and $v_c^{th}(R_i)$ are the observed (with a measurement error $\varepsilon_i$) and theoretically predicted values of the circular velocity at the radius $R_i$ of the i\,-\,th point, and the sum runs over the observed data points. The best fit is obtained by maximizing the likelihood ${\cal{L}}({\bf p})$, but it is worth stressing that, according to the Bayesian philosophy, the best estimate of the parameter $p_i$ is not the best fit one. On the contrary, one has to marginalize over the remaining parameters and look at the shape of the marginalized likelihood function defined as\,:

\begin{displaymath}
{\cal{L}}_i(p_i) \propto \int{{\cal{L}}({\bf p}) dp_1 \ldots dp_{i-1} dp_{i+1} \ldots dp_n}\,,
\end{displaymath}
with $n$ the total number of parameters. Actually, what we did was running a Monte Carlo Markov Chain (MCMC) code to efficiently explore the five dimensional parameter space and use of the histogram of the values for the parameter $p_i$ to estimate the mean, the median and the $68\%$ and $95\%$ confidence ranges. Note that, because of degeneracies among the parameters of the model, the best fit parameters ${\bf p}_{bf}$ may also differ from the maximum likelihood ones ${\bf p}_{ML}$, i.e. the set obtained by maximizing each of the marginalized likelihood functions.

It is mandatory to explain which are the parameters we constrain by our fitting procedure. Firstly, we scale the HL characteristic radii $(r_A, r_D)$ with respect to the disc scalelength $R_d$, and then skip to logarithmic units using $(\log{\eta_A}, \log{\eta_D})$ as model parameters in order to explore a wider range. As already quoted above, the halo model is assigned by three parameters which we choose to be $(\alpha, c_{vir}, V_{vir})$, while the virial mass is estimated a posteriori from the $V_{vir}$ value. Finally, although the total disc mass has been set from the beginning, it is worth noting that such an estimate has been obtained converting colors into a fiducial $M/L$ ratio $\Upsilon_{fid}$ assuming a diet Salpeter initial mass function (IMF) and a given recipe for the different ingredients entering the stellar population synthesis code. As a consequence, one can not exclude a priori that the actual $M/L$ ratio $\Upsilon$ does not equal the fiducial one so that we add $\Upsilon/\Upsilon_{fid}$ to the list of fitting parameters

\section{Results}

\begin{figure*}
\centering
\subfigure{\includegraphics[width=7.5cm]{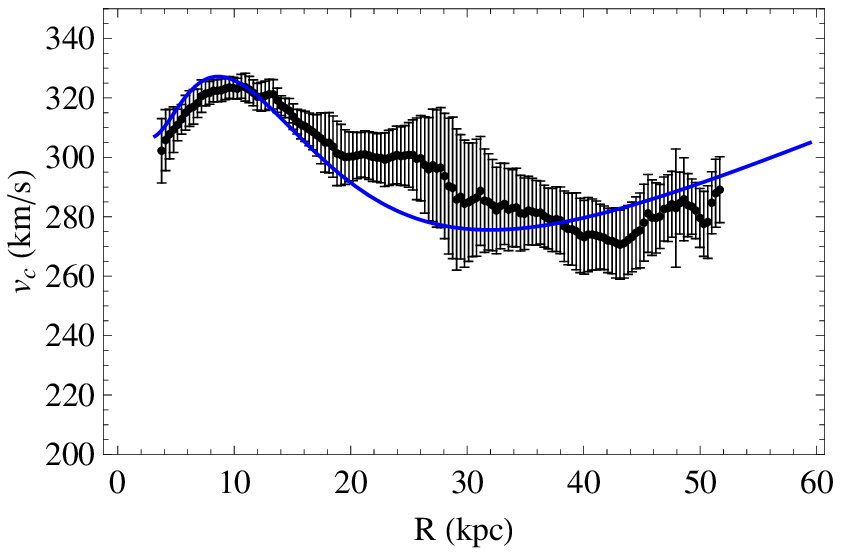}} \goodgap
\subfigure{\includegraphics[width=7.5cm]{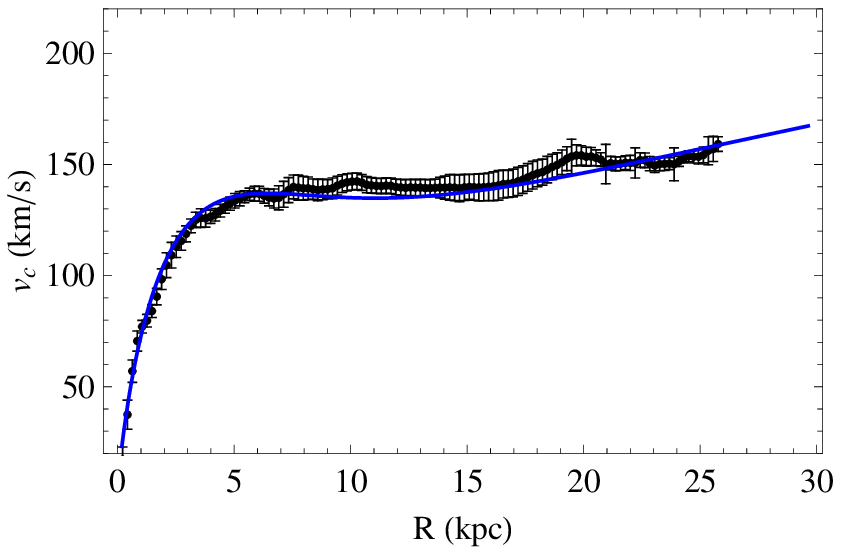}} \\
\subfigure{\includegraphics[width=7.5cm]{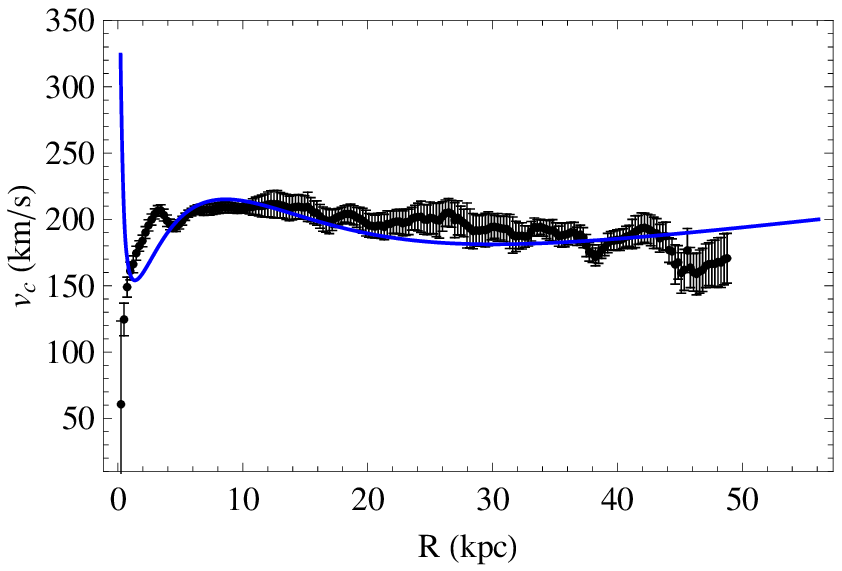}} \goodgap
\subfigure{\includegraphics[width=7.5cm]{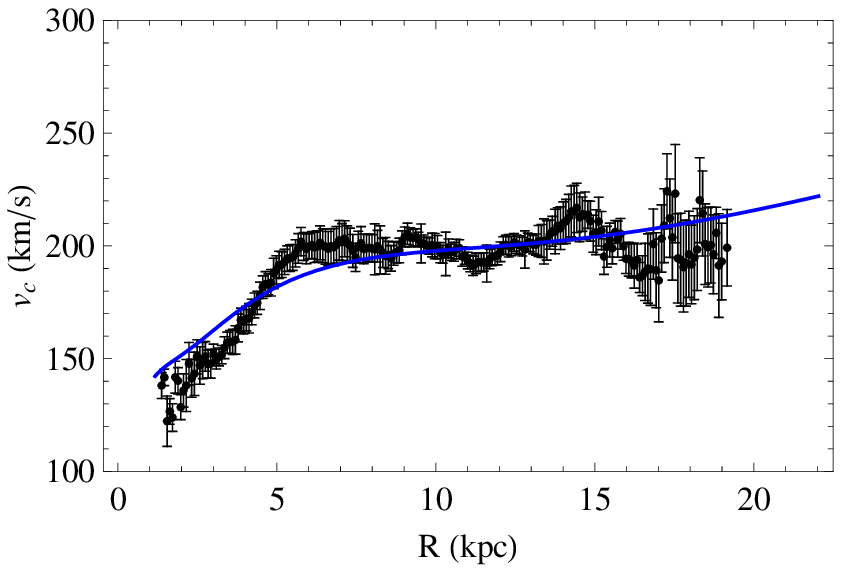}} \\
\caption{Best fit curves superimposed to the data for NGC 2841 (top left), NGC 3621 (top right), NGC 5055 (bottom left), NGC 6469 (bottom right). See Table 2 for the values of the best fit parameters.}
\label{fig: bf}
\end{figure*}

In order to constrain the parameters of the model, we have run, for each galaxy, three Markov Chains using the Gellman\,-\,Rubin (1992) criterium to check the convergence of the algorithm. The best fit solution and the constraints on the parameters are summarized in Table 2, while Fig.\,\ref{fig: bf} shows the best fit curve superimposed to the data.

As a preliminary remark, we remember the reader that the errors $\varepsilon_i$ on each data point are not Gaussian distributed, since they also take into account systematic misalignments between HI and H$\alpha$ measurements and other effects leading to a conservative overestimate of the true uncertainties. The final error on each point is, indeed, computed by adding in quadrature the dispersion around the best fit tilted ring model estimate and a pseudo $1 \sigma$ uncertainty due to the differences between the approaching and receding sides of the circular velocity profile (see \citealt{Things} for further details). Note that such procedure leads to larger errors for galaxies with residual non circular motions and/or warped HI distribution in the outer regions. As a consequence of the non Gaussian uncertainties, a value $\tilde{\chi}^2 = \chi^2/dof \simeq 1$ for the best fit model (with $dof = N - 6$ the number of degrees of freedom) could also be an evidence for an overestimate of the uncertainties rather than the outcome of the model closely matching the data.

That this is indeed the case may be easily understood noting that we get $\tilde{\chi}^2 = 0.92$ for NGC 2841 which is clearly not well fitted by the model. On the contrary, for the other systems, $\tilde{\chi}^2$ is a reasonably good indicator being $\tilde{\chi}^2 = 1.05$ for NGC 3621, $\tilde{\chi}^2 = 3.62$ for NGC 5055, $\tilde{\chi}^2 = 1.74$ for NGC 6469, in qualitative agreement with what one expects from the overall agreement between the model and the data. In order to get some insight on the agreement between the model prediction and the data, it is worth briefly discussing each galaxy on a case\,-\,by\,-\,case basis. \\

{\it NGC 2841.} This is the worst fitted galaxy in the sample with only the very inner regions well fitted by the data. While the data shows a decreasing trend, the best fit circular velocity profile starts diverging after an initial decrease. Notwithstanding the relatively small $\tilde{\chi}^2$ value, we can nevertheless safely argue that the modified potential is not able to fit the NGC 2841 rotation curve.

As a further evidence, we can also consider the marginalized constraints on the model parameters. First, we note that $\Upsilon/\Upsilon_{fid}$ takes quite large values. Since the diet Salpeter IMF used to infer $\Upsilon_{fid}$ provides the highest possible $M/L$ value compatible with the Tully\,-\,Fisher relation, it is quite hard to conceive a particular combination of stellar population properties able to give values of $\Upsilon/\Upsilon_{fid} \ge 1.25$ so that we can safely deem as unrealistic the $\Upsilon/\Upsilon_{fid}$ values needed to reconcile the model with the data.

Similarly, the concentration parameter takes extremely large values. It is worth noting, however, that what we are referring to as {\it concentration} is not the usually defined one. Indeed, one typically refers to $c_{-2} = R_{vir}/R_{-2}$ with $R_{-2}$ the radius where the logarithmic slope of the density profiles equals -2. For the gNFW model we are considering, it is $c_{-2} = c_{vir}/(2 - \alpha)$ so that, for the median $(\alpha, c_{vir})$ values in Table 2, we get $c_{-2} \sim 200$ which is still quite large if compared to the values in literature ($c_{-2} \sim 10 - 15$). Although previous estimates have been obtained in a Newtonian framework, it is unlikely that such strongly concentrated haloes are indeed realistic. It is therefore worth wondering what is driving the fit towards such large $c_{vir}$ values. To this end, let us first note that the estimated virial mass is given by $\log{M_{vir}} = 10.79_{-0.72}^{+0.64}$ which is typical for a spiral galaxy. Since $c_{vir} = R_{vir}/R_s$ and $R_{vir}$ takes typical values, we must conclude that it is $R_s$ to be unrealistically small.

In order to understand why such values are preferred, we note that the smaller is $R_s$, the earlier the regime $r/R_s >> 1$ is achieved and the halo density profile starts scaling as $r^{-3}$. As a consequence, for $R >> R_s$, the Newtonian contribution to the halo tends to quickly vanish and the outer regions circular velocity is mainly determined by the HL $r^2$ term. On the contrary, in the very inner regions, $R/R_s << 1$, the halo Newtonian $v_c(R)$ becomes quite small and its role is played by the HL $r^{-4}$ term. Indeed, the values of $(\log{\eta_A}, \log{\eta_D})$ are tailored in such a way to compensate a missing halo term in the outer and inner regions, respectively. Should $R_s$ not have been extremely small, the halo Newtonian term would contribute significantly in these two regions thus leading to a mismatch with the data. \\

{\it NGC 3621.} Contrary to NGC 2841, this galaxy is well fitted by the modified potential. However, a deeper look at the data shows that, while the inner regions are perfectly matched, the agreement between the theoretical curve and the data, although overall good, is getting worse as $R$ increases. Indeed, while the data suggests a flat behaviour, the best fit circular velocity starts increasing linearly as a consequence of the $r^2$ term in the HL potential leading to $v_c(R) \propto R$ for $R/R_d >> \eta_A$. Should one be able to probe $v_c(R)$ to larger $R$, the model will likely fail to fit the extended dataset leading to a situation similar to NGC 2841.

Table 2 shows that the concentration is still quite large so that the same qualitative discussion made for NGC 2841 also holds here. Moreover, the estimated virial mass, $\log{M_{vir}} = 9.78_{-1.03}^{+0.67}$, turns out to be smaller than the disc mass which can lead to serious problems on cosmological scales (see later for a discussion of this issue). \\

{\it NGC 5055.} The striking feature of this fit is the opposite behaviour of the theoretical and observational rotation curve in the very inner regions. Although such a strong disagreement for a best fit model could be surprising, it is worth noting that actually only two points are missed so that their weight in the overall $\chi^2$ is quite small compared to the other ones probing the intermediate and outer regions. What we have suggested for NGC 3621 clearly takes place for this galaxy. The rotation curve data stays roughly flat (or slowly decreasing at most), while the theoretical profile first decreases and then starts increasing linearly as consequence of the set in of the $r^2$ term in the HL modified potential.

Again, we find a very large concentration indicating an extremely small $R_s$ so that the qualitative discussion made for NGC 2841 holds for this case too. Moreover, the virial mass, $\log{M_{vir}} = 9.84_{-0.91}^{+0.99}$, is one order of magnitude smaller than the disc mass so that, contrary to the Newtonian case, the system is stars dominated everywhere. \\

{\it NGC 6469.} From an observational point of view, NGC 6469 can be considered a classical example of a galaxy with a rotation curve first increasing almost linearly and then setting in a long flat profile (although with some wiggles likely due to a clumpy gas distribution). On the contrary, the best fit curve poorly matches the rising part and shows a definitely increasing trend which does not follow the trend probed by the data. Moreover, we again find large concentrations and a virial mass, $\log{M_{vir}}= 9.92_{-1.25}^{+1.02}$, smaller than the disc mass.

\subsection{Discussion}

What the above results tell us is that our disks\,+\,halo model gives rise to a theoretical rotation curve which can not always be reconciled with the observed one. While the theoretical best fit circular velocity provide a satisfactory agreement with the data for NGC 3621 and NGC 5055 (excluding the very inner points), it traces the data for NGC 2841 and NGC 6469 quite poorly. Moreover, in all cases, there is marked trend indicating a rising rotation curve in the outer regions, whereas an extrapolation of the data suggests a flat profile. Although one could consider this latter argument questionable, it is worth stressing that, in order the fit to be acceptable, the price to pay is to have large concentrations and small halo masses. Being the rotation curves used of high quality and not affected by any systematic due to wrong inclination or distance determination, we can safely argue that the poor matches among the best fit theoretical curve and the measured one is a strong evidence of a failure of the modified HL potential.

Two further considerations strengthen this conclusion. First, we have qualitatively explained the large concentrations as a consequence of the need to reduce as much as possible the halo contribution in the very inner and very outer regions. For the median values of $\log{\eta_A}$ and $\log{\eta_D}$, the two corrective terms introduced by the HL modified potential boost the Newtonian circular velocity for $R/R_d >> \eta_A$ and $R/R_d << \eta_D$. Since the disc parameters are held fixed, the halo scaling quantities (mass and radius) have to be adjusted in order to not lead to $v_c(R)$ values larger than the observed ones. In particular, this drives the best fit model towards halo virial masses smaller than the disc ones. Considering the large error bars, one can at most have $\log{M_{vir}} \simeq \log{M_d}$, which, if extrapolated to the cosmological scales, would imply $\Omega_b \simeq \Omega_{CDM}$, quite at odds with what is expected from the expansion and growth of structures data\footnote{It is worth stressing that, for a spatially flat universe, the Friedmann equations for the HL gravity reduce to those for the concordance $\Lambda$CDM model so that one expects the same matter content in both theories in order to fit, e.g., the SNeIa Hubble diagram.}.

Table 2 shows that the $68\%$ confidence ranges of $(\log{\eta_A}, \log{\eta_D})$ for the different galaxies overlap quite well so that, although a larger sample is needed, one could argue that the scaled radii $(\eta_A, \eta_D)$ are universal quantities. Actually, such a result would represent a shortcoming for the consistency of the theory. Indeed, since $R_d$ is different from one galaxy to another, one should postulate a coupling between the properties of the galaxy and the coupling parameters entering the HL Lagrangian (and determining the $(r_A, r_D)$ radii) leading to a universal value of $(\eta_A, \eta_D)$. This would lead to a gravity theory depending on the particular system one is considering which is clearly at odds with the fundamental properties the gravity Lagrangian must have. A possible way out would be to invoke a non universality of the local MF, but it is hard to find a physically motivated recipe linking the disc scale radius $R_d$ and the MF in such a finely tuned way to make $(\eta_A, \eta_D)$ universal quantities.

Motivated by the above analysis, we can therefore complement the result in Paper I (where no halo was considered) concluding that the HL modified gravitational potential we have focused on is unable to reproduce the spiral galaxies rotation curve because of a too large boost to the circular velocity of both the visible and dark matter components.

\section{Conclusions}

Initially motivated by its attractive features from the point of view of quantum gravity, the HL proposal has soon become one of the most investigated theories of gravity alternative to GR. In \cite{PapI}, we have complemented recent works on its cosmological consequences by addressing its impact on the gravitational potential finding that, under some suitable conditions on the couplings constants entering its Lagrangian, it is possible to work out a modified gravitational potential made out of the Newtonian $1/r$ term corrected by the addition of two further terms scaling as $r^2$ and $r^{-4}$, respectively. We have here made a step further investigating the viability of this modified potential from the point of view of galaxy dynamics. To this end, we have tried fitting the rotation curve data of 4 spiral galaxies including both visible (bulge and disc) and dark matter components. As a result, we have found that the theoretical rotation curve may match the observed one only in the very inner regions, but the theory parameters $(\eta_A, \eta_D)$ can not be adjusted to fit also the outer regions. Moreover, the halo virial mass and concentration should take unreasonably odd values, the halo mass and characteristic radius being smaller than their disc counterparts. While one could look at such a low dark matter content as a good news, it is worth stressing that HL gravity still needs substantial dark matter on cosmological scales in order to provide the correct evolution for the scale factor. As such, lowering the galaxies dark matter content does not solve the missing mass problem, but simply turns it into the problem of understanding where dark matter should be present in order to fill the gap with the needed cosmological amount.

It is worth stressing that, although the above results have been obtained assuming a particular halo model, we expect that they are qualitatively model independent. As discussed above, lowering the halo mass and scale radius is needed in order to compensate the boost in circular velocity due to the $r^{-4}$ and $r^2$ terms of the HL modified potential. Since they dominate in the inner and outer regions, respectively, one has to reduce the total mass in these regions to not override the observed $v_c(R)$. Having fixed the stars content, the only way to achieve to do this is indeed to have a large concentration and a small virial mass. Changing the halo model will likely change the actual values of $(c_{vir}, M_{vir})$, but can not reduce the boost due to the HL corrective terms. As a consequence, we can safely argue that similar results should finally be obtained thus qualitatively confirming the analysis presented here.

As a final remark, we point out that our discussion suggests that also the Schwartzschild\,-\,de Sitter solution (which presents the $r^2$ term, but not the $r^{-4}$ one), obtained in the framework of SVW version of HL gravity, can hardly fit the galaxies rotation curves. We can consequently conclude that the analysis of galaxies dynamics represents a strong evidence against the reliability of \textit{any} gravitational potential obtained in this modified version of HL gravity. \\

{\it Acknowledgements.} VFC warmly thanks W.J.G. de Blok for making available the THINGS data in electronic format and for the many enlightening discussion on the galaxies properties. The authors also acknowledge an unknown referee for his/her comments which have helped improving the paper. VFC and MC are supported by Italian Space Agency (ASI) and Regione Piemonte and Universit\`{a} degli Studi di Torino, respectively.

\end{document}